\documentclass{article}
\usepackage{ijcai27}
\usepackage{times}
\usepackage{soul}
\usepackage{url}
\usepackage[hidelinks]{hyperref}
\usepackage[utf8]{inputenc}
\usepackage[small]{caption}
\usepackage{graphicx}
\usepackage{amsmath}
\usepackage{amsthm}
\usepackage{amssymb}
\newtheorem{proposition}{Proposition}
\usepackage{booktabs}
\usepackage{float}
\newcommand{\ncalAcc}{27}
\newcommand{\ncalDet}{75}
\newcommand{\ncalFab}{3.3}
\newcommand{\ncalIncons}{0.27}
\newcommand{\ncalLoad}{4.3}
\newcommand{\ncalTurn}{0.48}
\newcommand{\ncolNC}{48}

\newcommand{\ncolRC}{0.99}
\newcommand{\ncolRSC}{0.95}
\newcommand{\ncoreDownGapHi}{23}
\newcommand{\ncoreDownGapLo}{12}

\newcommand{\neOneBestK}{35}

\newcommand{\neOneBestRel}{84}
\newcommand{\neOneBestRelSc}{128}
\newcommand{\neOneDvFifty}{-14.8}
\newcommand{\neOneDvNinety}{-41.2}
\newcommand{\neOneDvTen}{-0.5}

\newcommand{\neThreeBig}{+0.173}
\newcommand{\neThreeBigSc}{+0.367}
\newcommand{\neThreeOut}{-0.064}
\newcommand{\neThreeSmall}{-0.054}

\newcommand{\nharmCcorr}{-0.002}
\newcommand{\nharmCturn}{4.4}

\newcommand{\nharmRatioHi}{8}
\newcommand{\nharmRatioLo}{6}
\newcommand{\nharmScorr}{0.20}
\newcommand{\nharmSturn}{0.7}
\newcommand{\niclrAccQH}{0.285}
\newcommand{\niclrAccQS}{0.316}
\newcommand{\niclrAccUnqH}{0.282}
\newcommand{\niclrAccUnqS}{0.287}
\newcommand{\niclrNonQual}{63}
\newcommand{\niclrRingQual}{77}

\newcommand{\nnWorlds}{6,220}

\newcommand{\nopenFlags}{324}

\newcommand{\nscFlip}{31}

\newcommand{\nshortFour}{30}
\newcommand{\nshortTwo}{46}

\newcommand{\nsurgeGapH}{0.9}

\newcommand{\nsurgeGapHLongSig}{0.5}

\newcommand{\nsurgeGapSC}{2.7}

\newcommand{\nsurgeGapSCLongExp}{3.7}
\newcommand{\nsurgeGapSCLongSig}{3.2}

\newcommand{\nsurgeQualEnd}{68}
\newcommand{\nsurgeQualMin}{55}
\newcommand{\nsurgeQualStart}{74}

\newcommand{\ntauDownFixed}{10}
\newcommand{\ntauDownMean}{61}
\newcommand{\ntauSFixed}{-1.5}
\newcommand{\ntauSMean}{-13.5}
\newcommand{\ntrimCut}{31}

\newcommand{\nallyAccept}{20}
\newcommand{\nallyFixedRule}{0.45}
\newcommand{\nallyKeepHi}{93}
\newcommand{\nallyKeepLo}{69}
\newcommand{\nallyLLM}{0.50}
\newcommand{\nallyMeanRule}{0.56}
\newcommand{\nallyN}{20}
\newcommand{\nallyNum}{0.33}
\newcommand{\nallyRenew}{58}
\newcommand{\ncapOneFiftyCFiveHund}{-6.5}
\newcommand{\ncapOneFiftyCHund}{-35.0}
\newcommand{\ncapOneFiftyCThou}{-2.7}
\newcommand{\ncapOneFiftyCTwoFifty}{-18.3}

\newcommand{\neThreeTen}{+0.034}
\newcommand{\nllmCallsAll}{140,757}
\newcommand{\nllmDSBoostHi}{100}
\newcommand{\nllmDSBoostLo}{88}
\newcommand{\nllmDSFabHi}{0.7}
\newcommand{\nllmDSFabLo}{0.2}
\newcommand{\nllmDSFallback}{0.05}

\newcommand{\nllmDSKept}{44}
\newcommand{\nllmDSKeptHi}{58}
\newcommand{\nllmDSKeptLo}{36}
\newcommand{\nllmDSNonFixed}{12}
\newcommand{\nllmDSNonMean}{20}

\newcommand{\nllmDSOwnNon}{12}
\newcommand{\nllmDSOwnRing}{39}
\newcommand{\nllmDSRingFixed}{75}
\newcommand{\nllmDSRingMean}{80}

\newcommand{\nllmDSSep}{10}

\newcommand{\nllmDSSkipCMean}{30}
\newcommand{\nllmDSSkipHFixed}{13}
\newcommand{\nllmDSSkipHMean}{28}

\newcommand{\nllmQWBoostHi}{100}
\newcommand{\nllmQWBoostLo}{83}
\newcommand{\nllmQWFabHi}{56.1}
\newcommand{\nllmQWFabLo}{51.3}

\newcommand{\nllmQWKept}{71}
\newcommand{\nllmQWKeptHi}{90}
\newcommand{\nllmQWKeptLo}{49}
\newcommand{\nllmQWNonFixed}{13}
\newcommand{\nllmQWNonMean}{11}

\newcommand{\nllmQWNonMinusSMean}{1.4}
\newcommand{\nllmQWOwnNon}{15}
\newcommand{\nllmQWOwnRing}{26}
\newcommand{\nllmQWRingFixed}{30}
\newcommand{\nllmQWRingMean}{32}

\newcommand{\nllmQWTrimHi}{55}
\newcommand{\nllmQWTrimLo}{50}

\newcommand{\nllmWorldsAll}{90}

\newcommand{\nnumTwinOwnNon}{17}
\newcommand{\nnumTwinOwnRing}{14}

\newcommand{\nnumWorldsAll}{6,660}

\newcommand{\noosBiasSC}{-0.028}
\newcommand{\noosRMSEC}{0.023}
\newcommand{\noosRMSESC}{0.030}
\newcommand{\noosRsqC}{0.96}

\newcommand{\npopCells}{12}
\newcommand{\npopMaxDiff}{0.8}

\newcommand{\npropBoundBidOneFive}{0.344}
\newcommand{\npropBoundBidOneTwenty}{0.865}

\newcommand{\npropHitBidOneFive}{0.345}
\newcommand{\npropHitBidOneTwenty}{0.868}

\newcommand{\npropHitRandTwenty}{0.0077}

\newcommand{\npropLowRandTwenty}{0.0076}
\newcommand{\npropPiMaxFive}{2.60}
\newcommand{\npropPiMaxTwenty}{0.55}

\newcommand{\nscaleupWorlds}{440}

\newcommand{\nsurgeGapDiffCISigFive}{0.5}

\newcommand{\nsurgeGapDiffExpFive}{2.6}
\newcommand{\nsurgeGapDiffExpTwo}{3.5}
\newcommand{\nsurgeGapDiffExpTwoFive}{3.2}
\newcommand{\nsurgeGapDiffSigFive}{1.4}
\newcommand{\nsurgeGapDiffSigTwo}{2.9}
\newcommand{\nsurgeGapDiffSigTwoFive}{2.3}
\newcommand{\ntkBidOneFive}{11.2}
\newcommand{\ntkBidOneTwenty}{42.7}
\newcommand{\ntkBidThreeFive}{29.2}
\newcommand{\ntkBidThreeTwenty}{71.0}
\newcommand{\ntkPredFive}{0.050}
\newcommand{\ntkPredTwenty}{0.240}
\newcommand{\ntkRandFive}{0.055}
\newcommand{\ntkRandTwenty}{0.243}

\newcommand{\fig}[2][\textwidth]{\includegraphics[width=#1]{figures/#2.pdf}}
\newcommand{\pp}{percentage points}

\newif\ifarxiv
\arxivtrue

\title{When Reviews Gate Access: Collusion, Sabotage and the Scale of Peer Review\\in the AI Submission Surge}

\author{
Yuchen Li$^{*}$\and
Xintong Li$^{*}$\and
Ken-Tye Yong$^{\dagger}$\And
Nguyen H. Tran$^{\dagger}$\\
\affiliations
University of Sydney\\
$^{*}$Equal contribution.\quad $^{\dagger}$Corresponding authors.
}

\begin{document}

\maketitle

\begin{abstract}
AI tools have pushed submissions to the largest machine-learning venues to tens of thousands per cycle. Venues now tie access to review through submission caps, reciprocal reviewing, reviewer eligibility earned by acceptances, and bidding-based assignment. A reviewer is then also the authors' competitor. \emph{Sabotage} lowers a competitor's score and \emph{collusion} raises an ally's. We run \nnumWorldsAll{} conferences of up to 10{,}000 agents calibrated to NeurIPS and ICLR, and \nllmWorldsAll{} conferences of language-model agents from two families. First, collusion changes who is accepted, sabotage erodes what scores say, and each makes the other pay. Second, we prove that under random assignment collusion's reach is bounded by one exposure index $E=Rm/A$, so growth protects review, while bidding sets a floor that no pool size lowers. Among 10{,}000 reviewers a 20-member ring wins \ntkRandTwenty\% of its papers through collusion under random assignment and \ntkBidThreeTwenty\% when 30\% of allies' bids are granted. Third, when agents choose allies, caps at inner social-brain layers keep collusion nearly harmless, and the distortion of the accepted list follows one curve in $E$ that predicts larger pools out of sample. Fourth, ring membership makes language-model agents sabotage outsiders beyond what the ring's goal rewards, in both families.
\end{abstract}

\section{Introduction}

Peer review at the large AI venues was built for honest, independent reviewers and a manageable stream of papers. Both assumptions are under strain. AI writing tools lower the cost of producing a paper, and the largest venues now receive tens of thousands of submissions per cycle. Venues have answered by tying access to the review system itself. ICLR 2027 limits each author to 20 submissions, requires authors of three or more submissions to review at least six papers, and counts only authors with accepted papers at major venues as qualified reviewers \cite{iclr2027}. Assignment at this scale relies on reviewer bids \cite{cui2026}.

These rules turn a review score into a currency. A low score lowers a competitor's chance of acceptance and, under eligibility rules, can keep the competitor out of the reviewer pool. A high score does the reverse for an ally. We call the first deviation \emph{sabotage} and the second \emph{collusion}. Collusion rings have already been reported at computer-science venues \cite{littman2021,shah2022}.

\begin{figure*}[t]
\centering\fig[0.86\textwidth]{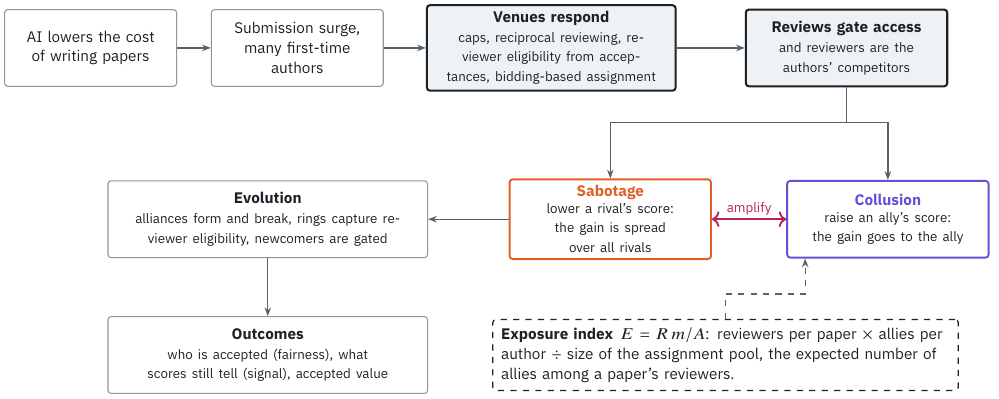}
\caption{\textbf{The question.} Rules adopted for the submission surge make review scores gate access, and reviewers become the authors' competitors. Sabotage and collusion then interact, shape how alliances and eligibility evolve, and decide who is accepted. The exposure index $E$ links the venue's design to collusion's reach.}
\label{fig:concept}
\end{figure*}

We ask four questions. Do sabotage and collusion interact? When does the size of a venue protect it against collusion? How do alliances evolve when agents choose them? How does reviewer eligibility shape who is accepted during the surge? We answer them with an agent-based conference calibrated to public statistics of NeurIPS and ICLR (Figure~\ref{fig:concept}), and we test the behavioural core with agents driven by language models. Our contributions are the following.

\begin{itemize}
\item \textbf{A testbed with two kinds of agents} (Sections~\ref{sec:model} and~\ref{sec:llm}). We build a multi-cycle conference calibrated to NeurIPS and ICLR, in which numerical agents optimise their acceptances under the new rules, and we run the same rules as a game played by language-model agents. The numerical model covers \nnumWorldsAll{} conferences of up to 10{,}000 agents, and the game \nllmWorldsAll{} conferences played by two model families.
\item \textbf{Two harms that amplify each other} (Section~\ref{sec:harms}). Collusion changes who is accepted and sabotage erodes what scores say about quality. We show that a ring multiplies the expected gain of each act of sabotage by its size at the same risk.
\item \textbf{An exposure law and a scale paradox} (Section~\ref{sec:scale}). We prove that under random assignment the chance that a paper meets an ally lies between $1-e^{-E}$ and $E=Rm/A$, so growth protects review, and that bidding sets a floor of $1-(1-\pi)^m$ that no pool size lowers. The floor gives a design rule for the grant rate of bids, and both results hold in a pool of 10{,}000 reviewers.
\item \textbf{Alliance evolution} (Section~\ref{sec:alliance}). Caps at the inner social-brain layers keep collusion nearly harmless. The distortion of the accepted list is one function of $E$ and predicts pools of 500 and 1{,}000 out of sample.
\item \textbf{An eligibility loop} (Section~\ref{sec:surge}). When only authors with acceptances may review, sabotage favours incumbents, and a longer eligibility window widens the newcomer gap only when sabotage is possible.
\item \textbf{Rings breed sabotage in language-model agents} (Section~\ref{sec:llm}). In two model families with five seeds each, ring members sabotage outsiders more than other reviewers, and \nllmDSKept\% (DeepSeek) and \nllmQWKept\% (Qwen) of the gap remain when the ring's joint goal is removed.
\end{itemize}

\section{Related Work}\label{sec:related}

\paragraph{Strategic referees in agent-based models.} A small share of referees who reject papers better than their own pushes accepted quality towards the average \cite{thurner2011}. Disagreement among referees can be used to control such cheating \cite{grimaldo2013}. Later models add competition, reciprocity and resource constraints \cite{cabota2013,squazzoni2013,bianchi2018}. These models have strategic downgrading but no collusion rings, no gated access and no growth.

\paragraph{Collusion rings and assignment.} Randomised assignment caps the probability that a reviewer receives a given paper \cite{jecmen2020}. Assignments without short review cycles are hard to compute but can be found in practice \cite{boehmer2022}. Rings can be detected from bids \cite{jecmen2024} or with a score inspired by the VCG mechanism \cite{rabinovitz2026}, and robust assignment scales to tens of thousands of papers \cite{cui2026}. Strategyproof assignment and tests for strategic reviewing address sabotage in one cycle \cite{xu2019,stelmakh2021}. These studies are static or cover one cycle. We model behaviour that responds to the rules over many cycles and show how the assignment rule sets collusion's reach through one index.

\paragraph{Collusion under the new rules.} Collusion between strangers through bidding is not individually rational in one shot \cite{xing2026bidding}, and credential-gated reciprocal review can be gamed \cite{xing2026governance}. CABAL simulates reviewers driven by language models with collusive bidding in one cycle and finds modest effects on the venue as a whole \cite{zhou2026cabal}. A position paper calls for agent-based stress tests of collusion before new mechanisms are deployed \cite{farooq2026}. We provide such a test with sabotage, repeated interaction, chosen alliances and growth.

\paragraph{Language-model agents.} AgentReview simulates reviewers, area chairs and authors driven by language models \cite{jin2024}. Language models show in-group solidarity and out-group hostility \cite{hu2025}, and persona agents favour their in-group over facts \cite{lei2026}. Our results agree with two earlier findings: small rings displace papers without lowering aggregate quality \cite{xing2026bidding,zhou2026cabal}, and language-model agents carry in-group bias \cite{hu2025,lei2026}. We add the cost of that bias to other agents and measure its size against a payoff model.

\section{Model}\label{sec:model}

\subsection{A Calibrated Conference}
Every agent is both an author and a reviewer in one research area. The baseline world has 500 agents in five areas of 100 and runs for 20 rounds. A round is one review cycle of the ecosystem of top AI venues, so 20 rounds span about two to three years. In each round an agent receives a research budget and a few research ideas. For each idea it skips, writes an honest paper with more or less effort, or fabricates a result. Effort raises quality with diminishing returns. A fabricated paper looks stronger, but each reviewer detects it with some probability, and a detected author is barred for two rounds. Each paper receives $R=4$ reviewers from its area, who see its quality with Gaussian noise and report a score on the ICLR scale (1, 3, 5, 6, 8, 10). In each area the top 27\% of papers by mean score are accepted.

The honest world is calibrated before any deviation is switched on. It gives an acceptance rate of \ncalAcc\% and \ncalLoad{} reviews per reviewer. In the NeurIPS 2021 consistency experiment 23\% of decisions changed under an independent committee and about half of the accepted list turned over \cite{beygelzimer2023}. The model gives \ncalIncons{} and \ncalTurn. Fabricated papers make up \ncalFab\% of submissions and \ncalDet\% of them are caught, in line with survey estimates \cite{fanelli2009}.

\subsection{Two Deviations}
A reviewer who sabotages reports a score two levels below the honest one. A reviewer who colludes gives an ally's paper the top score. We compare four worlds: honest (H), sabotage allowed (S), collusion inside rings (C), and both (SC). In SC, ring members maximise the acceptances of the whole ring. A review that differs from the honest score is audited with probability 1\%, and an audited reviewer cannot submit for two rounds. Rings are fixed in advance or formed by the agents in every round (Section~\ref{sec:alliance}). By default rings of five hold 20\% of each area.

\subsection{Rules, Resources and Assignment}
Under \emph{Fixed} every agent may submit three papers per round. Under \emph{Mean} the cap, from one to five, rises with the agent's smoothed average past score, and under \emph{Tail} it falls with the share of low scores received. Under an \emph{ICLR-27}-like rule only qualified agents review and may submit three papers, and others may submit one. An agent qualifies with an accepted paper and keeps the status forever or for a window of $W$ rounds. Budgets are equal, or the richest 20\% receive four times the budget of the rest. Assignment is random within an area with balanced load. Under bidding, an ally's bid to review a ring member's paper is granted with probability $\pi$ before the remaining slots are filled at random.

\subsection{Decision Makers}
Numerical agents are forward-looking. Each chooses the option with the highest expected number of acceptances, now and later through the rule's effect on its future cap, minus the expected cost of a ban. Choices carry a small logit noise $\tau$. One act of sabotage spreads its gain over all competitors in the area, so its value is of the order of this noise, and we take sabotage magnitudes from noise-free runs ($\tau=0$). We also ran agents that imitate successful peers, and agents driven by a language model (Section~\ref{sec:llm}).

\subsection{Growth}
To mimic the surge, the population grows from 500 to 2{,}000 over 20 rounds, on a logistic curve in the main arm and by a constant 7.6\% per round in a second arm. Newcomers start without allies and without eligibility, and their ability has the same distribution as that of incumbents. At the start 70\% of agents are eligible, chosen by ability with noise.

\subsection{Design and Outcomes}
Every random draw is keyed by seed, purpose, round and agent, so worlds with the same seed share agents, ideas, review noise and rings, and every contrast between behaviours is paired. The main grids run 10 seeds (\nnWorlds{} conferences in 11 grids). A scale-up of \nscaleupWorlds{} further conferences with five seeds raises the population to 5{,}000, the reviewer pool to 10{,}000, the pools for chosen alliances to 1{,}000, and the surge to 5{,}000. Scaling the population to 2{,}500 and 5{,}000 agents in areas of 100 changes no result. Across \npopCells{} combinations of behaviour, rule and resources, accepted value relative to H moves by at most \npopMaxDiff{} points. \emph{Accepted value} is the summed true value of accepted papers relative to H with the same seed. \emph{Overlap} is the share of the list that honest scores would accept that is in fact accepted. The \emph{ring advantage} is the acceptances per round of ring members minus those of others.

\section{Results with Numerical Agents}\label{sec:results}

\subsection{Two Harms That Amplify Each Other}\label{sec:harms}
\begin{figure*}[t]
\centering\fig[0.70\textwidth]{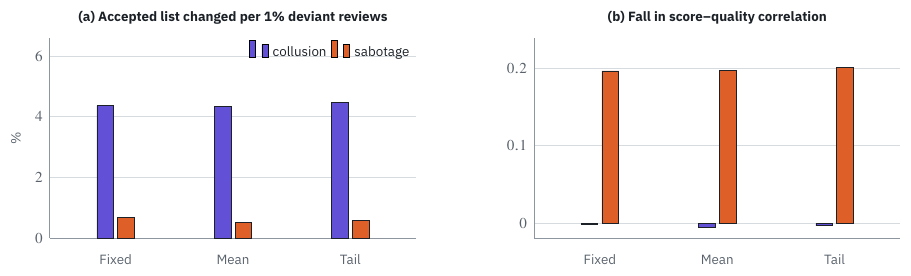}
\caption{\textbf{Collusion moves the list, sabotage blurs the signal.} Core grid, 500 agents, 20 rounds, 10 seeds. (a) Share of the accepted list that changes per 1\% of deviant reviews. (b) Fall in the correlation between mean score and true value.}
\label{fig:harms}
\end{figure*}

Collusion and sabotage harm different things (Figure~\ref{fig:harms}). Per 1\% of deviant reviews, collusion changes \nharmCturn\% of the accepted list and sabotage \nharmSturn\% under Fixed, a ratio of \nharmRatioLo{} to \nharmRatioHi{} across rules. Sabotage removes \nharmScorr{} of the correlation between scores and quality, while collusion leaves it intact ($\nharmCcorr$). Collusion is a problem of \emph{who} is accepted, and sabotage of \emph{what scores say}.

The two amplify each other. In all 12 combinations of rule, resources and budget feedback, rings that may also sabotage gain more than rings that only collude, and ring reviewers push outsiders down \ncoreDownGapLo{} to \ncoreDownGapHi{} \pp{} more often than other reviewers (Appendix~\ref{app:numerical}). The mechanism is simple. Sabotage that pushes an accepted paper below the cutoff frees one seat, and the seat goes to a rejected paper that the saboteur cannot choose.

\begin{proposition}[Rings multiply the gain from sabotage]\label{prop:ring}
Let $Q$ be the rejected papers that may take a freed seat, and let the seat go to each of them with equal probability. A reviewer with $q$ papers in $Q$ gains $q/|Q|$ expected acceptances from one successful act of sabotage. A member of a ring of $k$ agents with $q$ papers each in $Q$, who counts the ring's acceptances, gains $kq/|Q|$. The audit risk is the same for both.
\end{proposition}

A ring thus makes each act of sabotage worth $k$ times as much at the same risk (proofs in Appendix~\ref{app:proofs}). Rules that tie future caps to past scores multiply sabotage. Without decision noise reviewers sabotage \ntauDownFixed\% of the time under Fixed and \ntauDownMean\% under Mean, and accepted value changes by $\ntauSFixed$\% and $\ntauSMean$\%.

\subsection{The Exposure Law and the Scale Paradox}\label{sec:scale}
Consider a paper whose author has $m$ allies among the $A-1$ other reviewers of its pool, and let $E=Rm/(A-1)$, the expected number of allies among its $R$ reviewers. The paper is \emph{hit} if at least one reviewer is an ally.

\begin{proposition}[Exposure law]\label{prop:exposure}
If the $R$ reviewers are drawn at random without replacement, then
\begin{gather}
P(\text{hit}) = 1-\binom{A-1-m}{R}\Big/\binom{A-1}{R},\label{eq:hit}\\
1-e^{-E}\;\le\; P(\text{hit})\;\le\; E.\label{eq:bounds}
\end{gather}
\end{proposition}

A boost changes the decision only for papers near the cutoff, about \nscFlip\% of boosted papers. Collusion's \emph{success rate}, the share of a ring's papers accepted because of a boost, is $P(\text{hit})$ times this flip rate $\phi$. The design variables $R$, $m$ and $A$ therefore act only through $E$, and success is at most $\phi E$, which falls as $1/A$ when $R$ and $m$ are fixed.

\begin{figure*}[t]
\centering\fig[0.88\textwidth]{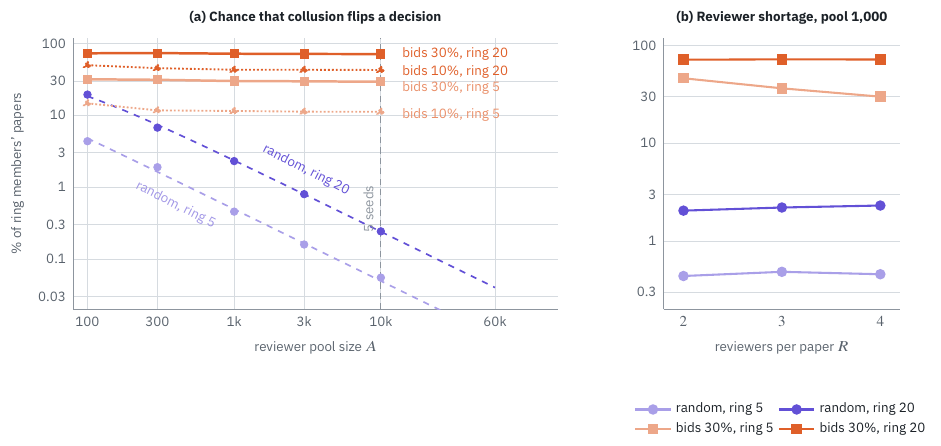}
\caption{\textbf{Growth protects only random assignment.} Rings of 5 and 20 in one pool, 10 rounds. Pools of 100 to 3{,}000 have 10 seeds, the pool of 10{,}000 five. (a) Share of ring members' papers accepted because of a boost. Dots: random assignment. Dashed: Equation~\eqref{eq:hit} times the flip rate of the smaller pools. Squares and triangles: allies' bids granted with probability 30\% or 10\%. (b) Reviewer shortage in a pool of 1{,}000.}
\label{fig:scale}
\end{figure*}

Simulated hit rates match Equation~\eqref{eq:hit} within 0.7 \pp{} for pools of 100 to 3{,}000 and rings of 5 and 20 (Figure~\ref{fig:scale}a). In the scale-up pool of 10{,}000 the hit rate of a 20-ring is \npropHitRandTwenty, with both bounds at \npropLowRandTwenty, and the ring succeeds with \ntkRandTwenty\% of its papers against a prediction of \ntkPredTwenty\% from the flip rate of the smaller pools (\ntkRandFive\% against \ntkPredFive\% for a 5-ring). Growth protects random assignment.

\begin{proposition}[Bidding floor]\label{prop:bid}
If each ally's bid to review the paper is granted independently with probability $\pi$ before the remaining seats are filled at random, then $P(\text{hit})\ge 1-(1-\pi)^m$ for every pool size $A$.
\end{proposition}

Bidding therefore removes the protection. Among 10{,}000 reviewers with $\pi=10\%$, the hit rate is \npropHitBidOneTwenty{} for a 20-ring against a floor of \npropBoundBidOneTwenty, and \npropHitBidOneFive{} for a 5-ring against \npropBoundBidOneFive. Success is flat in $A$, at \ntkBidOneFive\% for a 5-ring and \ntkBidOneTwenty\% for a 20-ring, and at \ntkBidThreeFive\% and \ntkBidThreeTwenty\% when $\pi=30\%$. The floor gives a design rule: keeping a ring's hit rate below $x$ requires $\pi\le 1-(1-x)^{1/m}$. For $x=10\%$ the grant rate must stay below \npropPiMaxTwenty\% for a 20-ring and \npropPiMaxFive\% for a 5-ring, a bound that per-pair caps on assignment probability \cite{jecmen2020} can enforce. Fewer reviewers per paper make each boost more decisive: with bidding, cutting $R$ from 4 to 2 raises a 5-ring's success from \nshortFour\% to \nshortTwo\% (Figure~\ref{fig:scale}b). A trimmed mean that drops the highest and lowest report removes \ntrimCut\% of a ring's advantage in pools of 50.

\subsection{Alliance Evolution}\label{sec:alliance}

\paragraph{Fixed rings.} With one ring in an area of 100, each member's gain rises with ring size and peaks at about \neOneBestK{} members (+\neOneBestRel\% acceptances per round, +\neOneBestRelSc\% with sabotage). The social cost grows much faster than linearly, from $\neOneDvTen$\% of accepted value at 10 members to $\neOneDvFifty$\% at 50 and $\neOneDvNinety$\% at 90. When rings of 2, 3, 5, 10 and 20 coexist, members of rings of up to five end up worse off than if no one colluded. A member of the 10-ring gains $\neThreeTen$ acceptances per round and a member of the 20-ring $\neThreeBig$. Big rings eat small ones (Appendix~\ref{app:numerical}).

\begin{figure*}[t]
\centering\fig[0.96\textwidth]{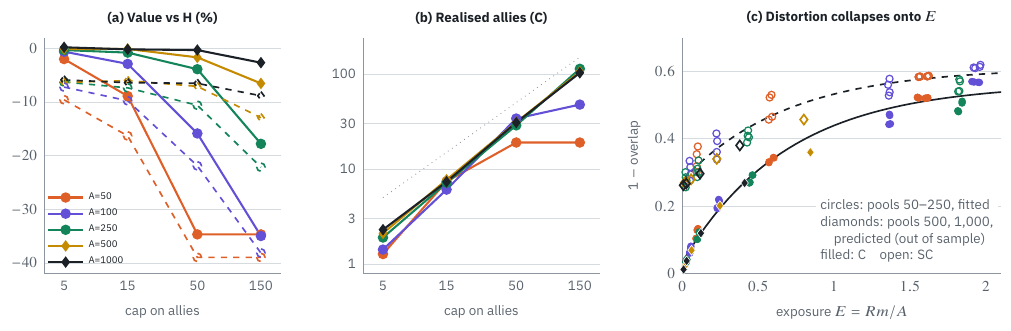}
\caption{\textbf{Alliance caps at the social-brain layers.} Agents choose allies every round, detection on. Pools of 50 to 250 have 10 seeds, pools of 500 and 1{,}000 five. (a) Accepted value relative to H as the cap grows (solid C, dashed SC). (b) Realised allies per agent, with the cap dotted. (c) Distortion of the accepted list against realised exposure. Circles are the cells used for the saturating fits (filled and solid: C, open and dashed: SC). Diamonds are the larger pools, predicted out of sample.}
\label{fig:dunbar}
\end{figure*}

\paragraph{Chosen alliances.} Here every agent proposes ties to the partners it values most in every round, recipients accept or reject, and agents above their cap drop their least valuable ties. A tie is valued by the acceptances it adds minus the expected cost of audits and detection. A pair that boosts each other twice within five rounds is caught with probability 0.3. We cap allies at the layers of the social brain, about 5, 15, 50 and 150 relationships \cite{dunbar1998,zhou2005}. Inner caps keep the network sparse and collusion nearly harmless (Figure~\ref{fig:dunbar}a). In areas of 100 under Fixed with equal budgets, a cap of 5 gives 1.4 allies per agent and changes accepted value by $-0.6\%$. A cap of 50 gives 34 allies and a 16\% loss (22\% with sabotage). Without a cap agents ally with half of their area and value falls by 35\%. Detection flags about \nopenFlags{} pairs per run but does not restore value.

\paragraph{The exposure collapse.} The cap alone does not set the damage. At a cap of 150 agents hold about 100 allies in pools of 250 to 1{,}000, yet accepted value changes by $\ncapOneFiftyCHund$\% at 100 reviewers and by only $\ncapOneFiftyCTwoFifty$\%, $\ncapOneFiftyCFiveHund$\% and $\ncapOneFiftyCThou$\% at 250, 500 and 1{,}000. One index governs the outcome. Across \ncolNC{} combinations of cap, pool (50 to 250), rule and resources, one minus the overlap is a saturating function of the realised exposure $E=Rm/(A-1)$ ($R^2=\ncolRC$ for C and \ncolRSC{} for SC, Figure~\ref{fig:dunbar}c). Fitted on these pools, the curve predicts the scale-up pools of 500 and 1{,}000 out of sample with $R^2=\noosRsqC$ and an RMSE of \noosRMSEC{} for C. For SC the RMSE is \noosRMSESC.

\subsection{The Eligibility Loop under the Surge}\label{sec:surge}
Under the ICLR-27-like rule, sabotage favours agents who start eligible. With sabotage they are accepted \niclrAccQS{} times per round against \niclrAccUnqS{} for the rest, while the two are nearly equal in H (\niclrAccQH{} and \niclrAccUnqH). Only eligible agents review, so sabotage is in their hands. Rings keep their eligibility more often than others (\niclrRingQual\% against \niclrNonQual\% with a 5-round window under SC).

\begin{figure*}[t]
\centering\fig[0.92\textwidth]{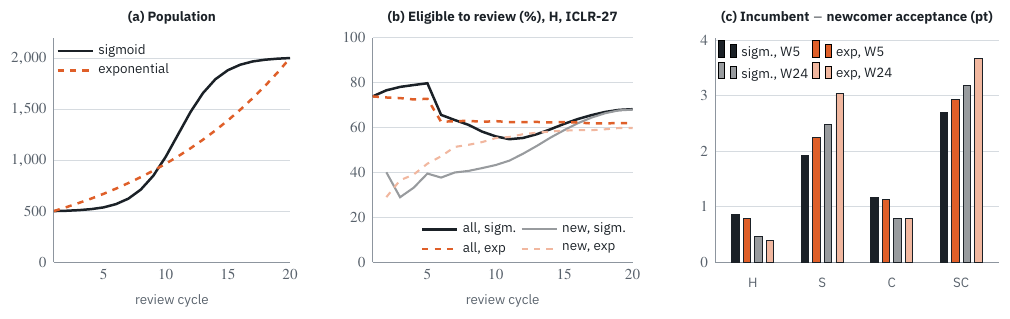}
\caption{\textbf{The surge.} Population 500 to 2{,}000 under logistic (sigm.) and exponential (exp) growth, ICLR-27-like rule, equal budgets, 10 seeds. (a) Population. (b) Eligible share of all agents and of newcomers in H. The step at round 6 is the end of the initial 5-round eligibility. (c) Incumbent minus newcomer acceptance rate per paper, with eligibility windows of 5 and 24 rounds.}
\label{fig:surge}
\end{figure*}

As the population grows from 500 to 2{,}000, the eligible share falls from \nsurgeQualStart\% to \nsurgeQualMin\% at the peak of the wave and recovers to \nsurgeQualEnd\% (Figure~\ref{fig:surge}b). Incumbents lead newcomers in acceptance rate per paper by \nsurgeGapH{} \pp{} in H and by \nsurgeGapSC{} in SC. A 24-round eligibility window, closer to real rules, shrinks the honest gap to \nsurgeGapHLongSig{} points but widens the gap in SC to \nsurgeGapSCLongSig{} (\nsurgeGapSCLongExp{} under exponential growth). A longer window shields incumbents only when sabotage is possible. In larger surges the effect of sabotage on the gap (SC minus H, 24-round window, five seeds) narrows but stays positive: \nsurgeGapDiffSigTwo{}, \nsurgeGapDiffSigTwoFive{} and \nsurgeGapDiffSigFive{} points ($\pm$\nsurgeGapDiffCISigFive{} at 5{,}000) for final populations of 2{,}000, 2{,}500 and 5{,}000 under logistic growth, and \nsurgeGapDiffExpTwo{}, \nsurgeGapDiffExpTwoFive{} and \nsurgeGapDiffExpFive{} under exponential growth.

\section{Language-Model Agents}\label{sec:llm}

\subsection{Setup}
We replaced the numerical decision rule by calls to a language model. Fifty agents in one area play 10 rounds with two rings of five. Each decision is one independent call that states the rules, the payoffs and the agent's own forecasts and asks for a JSON reply (Appendix~\ref{app:llm}). Authors choose which ideas to submit and how. A reviewer gets a sabotage option on rivals' papers in S and SC and an ally option on ring members' papers in C and SC. We used two families. DeepSeek V4.1 Flash (low reasoning effort) ran all eight combinations of H, S, C, SC with Fixed and Mean, an own-goal ablation, and four worlds in which agents choose their allies. Qwen3.7-Flash (thinking budget of 1{,}024 tokens) ran S and SC with both rules and the ablation. Every world ran with seeds 1 to 5, giving \nllmWorldsAll{} worlds and \nllmCallsAll{} calls.

\subsection{Rings Breed Sabotage beyond Their Incentive}
\begin{figure*}[t]
\centering\fig[0.94\textwidth]{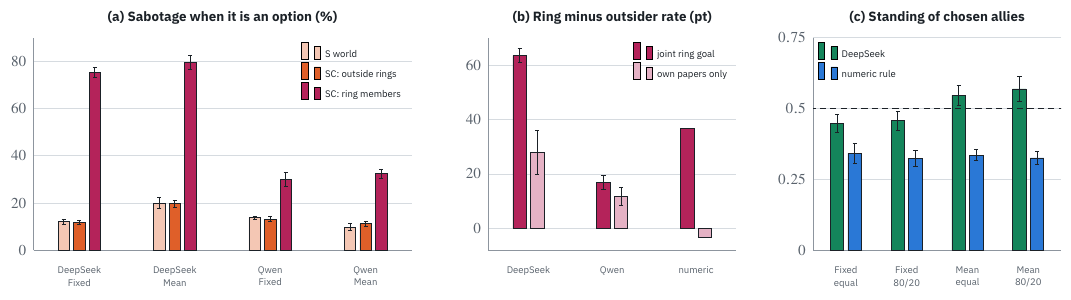}
\caption{\textbf{Language-model agents, two families, seeds 1 to 5.} Bars are means with 95\% intervals over seeds. (a) Share of sabotage options taken by reviewers in S worlds, and by reviewers outside and inside rings in SC worlds. (b) Ring members' minus other reviewers' sabotage rate in SC under Fixed, with the ring's joint goal or with each member's own papers only. The numerical agent is noise-free and averaged over 30 seeds. (c) DeepSeek agents choosing allies: percentile of chosen partners in past acceptances, against the numerical rule (0.5 is random).}
\label{fig:llm}
\end{figure*}

In both families ring members sabotage outsiders far more often than other reviewers (Figure~\ref{fig:llm}a). Under Fixed and Mean, DeepSeek ring members sabotage \nllmDSRingFixed\% and \nllmDSRingMean\% of their chances, against \nllmDSNonFixed\% and \nllmDSNonMean\% for reviewers outside rings. For Qwen the rates are \nllmQWRingFixed\% and \nllmQWRingMean\% against \nllmQWNonFixed\% and \nllmQWNonMean\%. Reviewers outside rings in SC sabotage as often as reviewers in S, within \nllmQWNonMinusSMean{} points, so the rise is tied to membership. In all \nllmDSSep{} DeepSeek SC worlds every ring member sabotaged more often than every other reviewer (counting reviewers with at least five chances).

A ring that maximises its joint acceptances has a reason to sabotage. To separate this goal from membership, we told ring members in SC under Fixed to count their own papers only (Figure~\ref{fig:llm}b). The gap between members and others shrinks but stays: DeepSeek members sabotage \nllmDSOwnRing\% against \nllmDSOwnNon\%, keeping \nllmDSKept\% of the gap (\nllmDSKeptLo--\nllmDSKeptHi\% across seeds), and Qwen members \nllmQWOwnRing\% against \nllmQWOwnNon\%, keeping \nllmQWKept\% (\nllmQWKeptLo--\nllmQWKeptHi\%). The noise-free numerical agent with the same goal shows no ring effect (\nnumTwinOwnRing\% against \nnumTwinOwnNon\%). This matches the in-group bias of language models \cite{hu2025,lei2026}, here with a cost to other agents. Members also keep their pacts, boosting allies' papers in \nllmDSBoostLo--\nllmDSBoostHi\% of chances for DeepSeek and \nllmQWBoostLo--\nllmQWBoostHi\% for Qwen.

\subsection{Partners Chosen by Standing}
When DeepSeek agents choose their allies, they pick partners at the median of past acceptances (percentile \nallyLLM, \nallyMeanRule{} under Mean and \nallyFixedRule{} under Fixed). The numerical rule picks low-volume partners (\nallyNum), below the language-model choice in all \nallyN{} worlds (Figure~\ref{fig:llm}c). Agents accept \nallyAccept\% of proposals, renew \nallyRenew\% of ties, and honour \nallyKeepLo--\nallyKeepHi\% of their pacts in review. Who benefits from collusion therefore depends on how partners are chosen.

\subsection{Families Differ as Authors}
The two families write very differently. DeepSeek authors almost never fabricate (\nllmDSFabLo--\nllmDSFabHi\% of papers). In H and C under Mean, \nllmDSSkipHMean--\nllmDSSkipCMean\% of their decisions submit nothing, which protects the average score, against \nllmDSSkipHFixed\% under Fixed. Qwen authors fabricate \nllmQWFabLo--\nllmQWFabHi\% of their papers. The ring effect on sabotage holds in both, so it does not depend on how agents write.

\section{Discussion}\label{sec:discussion}

\paragraph{Implications for venue design.} First, scale protects review only if assignment keeps the probability that a given pair meets low. Once allies can bid for each other's papers, a larger pool no longer dilutes collusion, and Proposition~\ref{prop:bid} sets the number: a 20-member ring stays below a 10\% hit rate only if allies' bids are granted less than \npropPiMaxTwenty\% of the time. Second, robust aggregation is cheap. A trimmed mean removes about a third of a ring's advantage, whereas adding reviewers does not help once sabotage is present. Third, tying the right to review to past acceptances puts sabotage in the hands of incumbents, and a longer eligibility window widens the newcomer gap only when sabotage occurs. Fourth, stress tests that use language-model agents should treat group membership as a variable. Membership alone raised sabotage beyond what payoffs explain, in both families.

\paragraph{Scope.} The model is calibrated on the honest aggregates of real venues, and its results identify the mechanisms and thresholds that venue designers can act on.

\section{Conclusion}
Rules that make review scores gate access turn reviewers into competitors. Collusion and sabotage then reinforce each other, bidding defeats the protection that growth gives random assignment, and eligibility tied to acceptances favours incumbents. One exposure index summarises collusion's reach across pools from 50 to 10{,}000. Agents driven by language models add in-group sabotage that a payoff model does not predict. Future work will derive the tipping point of alliance formation from $E$, price each policy lever against matching quality (including the matching benefit of bidding), calibrate scores on public OpenReview data, extend the game to more model families and wordings, and embed language-model agents in populations of thousands.

\bibliographystyle{named}
\bibliography{refs}

\ifarxiv
\clearpage
\appendix
\onecolumn
\begin{center}
{\Large\bfseries Technical Appendix}
\end{center}

\section{Proofs}\label{app:proofs}

\paragraph{Proposition~\ref{prop:ring}.} Condition on the act of sabotage succeeding, so that one seat is freed. By assumption the seat goes to each paper in $Q$ with probability $1/|Q|$, so the expected number of the reviewer's own papers that gain acceptance is $q/|Q|$. The objective of a ring member counts the papers in $Q$ of all $k$ members, $kq$ papers in total, so by linearity of expectation the expected gain is $kq/|Q|$. The audit probability and its penalty depend only on the reported score, not on the reviewer's objective, so the expected cost is the same. \hfill$\square$

\paragraph{Proposition~\ref{prop:exposure}.} The $R$ reviewers form a uniformly random $R$-subset of the $A-1$ other members of the pool. In the simulator each round shuffles the pool with a uniformly random permutation and gives each paper the next $R$ members after a uniformly random starting point, skipping the author, so for every paper the set of reviewers is uniform over the $R$-subsets of the other members. The number $N$ of allies among them is hypergeometric, and $P(N=0)=\binom{A-1-m}{R}/\binom{A-1}{R}$, which gives Equation~\eqref{eq:hit}. Written as a product,
\[
P(N=0)=\prod_{i=0}^{R-1}\frac{A-1-m-i}{A-1-i}=\prod_{i=0}^{R-1}\Bigl(1-\frac{m}{A-1-i}\Bigr)\le\Bigl(1-\frac{m}{A-1}\Bigr)^{R}\le e^{-Rm/(A-1)}=e^{-E},
\]
because each factor is at most $1-m/(A-1)$ and $1-x\le e^{-x}$. Hence $P(\text{hit})\ge 1-e^{-E}$. For the upper bound, each of the $R$ seats holds an ally with probability $m/(A-1)$, so $\mathbb{E}[N]=E$, and $P(\text{hit})=P(N\ge 1)\le \mathbb{E}[N]=E$ by Markov's inequality. \hfill$\square$

\paragraph{Proposition~\ref{prop:bid}.} The simulator processes the author's $m$ allies in random order before any random seat is filled, and grants each bid with probability $\pi$, independently. If at least one bid is granted, the first granted ally takes a seat, because no seat is filled before the bids are processed and $R\ge 1$. Hence $P(\text{hit})\ge P(\text{at least one bid granted})=1-(1-\pi)^m$, which does not involve $A$. A hit rate of at most $x$ therefore requires $1-(1-\pi)^m\le x$, that is $\pi\le 1-(1-x)^{1/m}$. \hfill$\square$

\section{Additional Results with Numerical Agents}\label{app:numerical}

\begin{figure}[H]
\centering\includegraphics[width=0.92\textwidth]{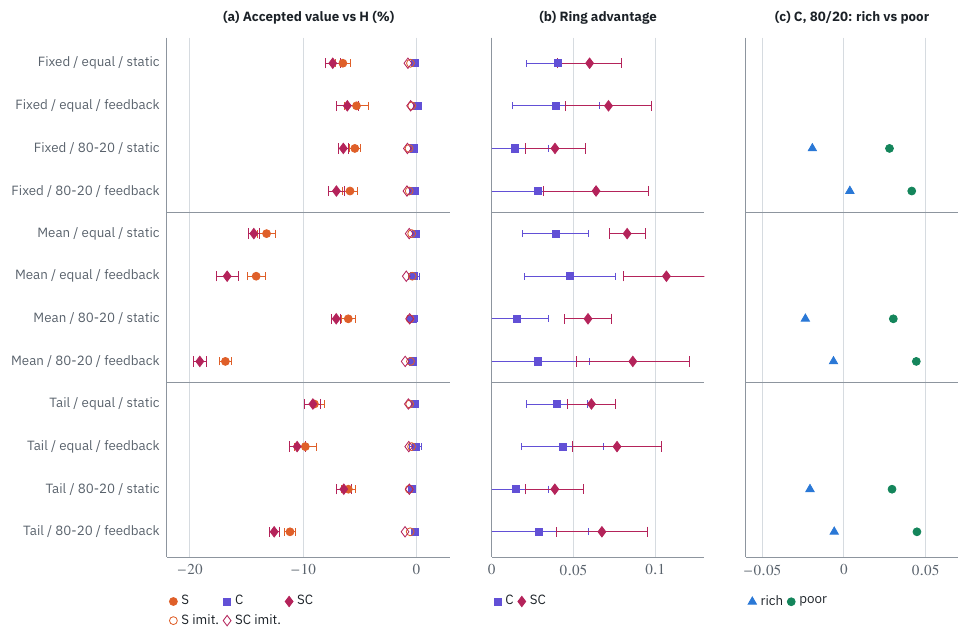}
\caption{\textbf{Core grid.} Behaviour $\times$ rule $\times$ resources (equal or 80/20) $\times$ budget feedback (static or feedback), 500 agents, 20 rounds, 10 seeds. (a) Accepted value relative to H with the same seed. Open markers are imitating agents. These runs use decision noise $\tau=0.01$, which inflates the numerical sabotage level. Without noise, sabotage changes accepted value by $\ntauSFixed$\% (Fixed) and $\ntauSMean$\% (Mean). (b) Ring advantage. (c) Under 80/20 budgets, the ring advantage among rich and among poor agents (C). Bars are 95\% intervals.}
\label{fig:core}
\end{figure}

\begin{figure}[H]
\centering\includegraphics[width=0.92\textwidth]{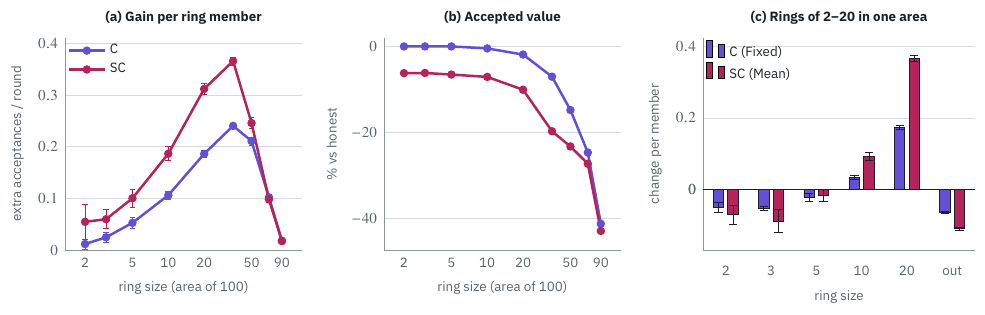}
\caption{\textbf{Ring size.} (a) One ring per area of 100: extra acceptances per round of each member against the same agent in H. (b) Accepted value relative to H. (c) Rings of 2, 3, 5, 10 and 20 in one area. Change per member against H for C under Fixed and SC under Mean. ``out'' marks agents in no ring. A member of a 3-ring changes by $\neThreeSmall$ acceptances per round, close to an outsider ($\neThreeOut$), while a member of the 20-ring gains $\neThreeBig$ ($\neThreeBigSc$ with sabotage under Mean). 10 seeds, 95\% intervals.}
\label{fig:ringsize}
\end{figure}

\begin{figure}[H]
\centering\includegraphics[width=0.7\textwidth]{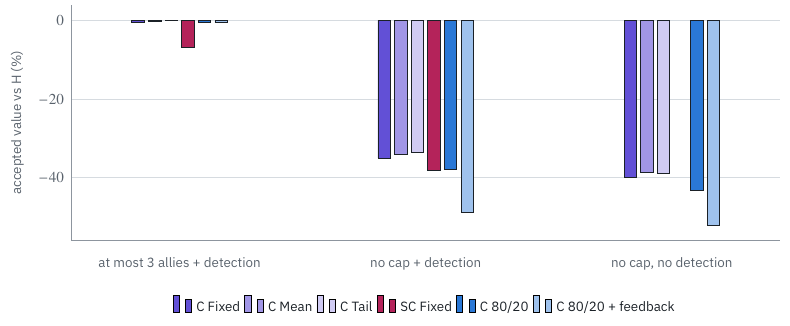}
\caption{\textbf{Free alliance formation.} Area of 100, no cap, three regimes, 10 seeds. Accepted value relative to H with the same seed. The rule (Fixed, Mean, Tail) barely matters. Unequal budgets with feedback deepen the loss to 49\%. Without detection agents hold 74 allies and value falls by 40\%.}
\label{fig:alliance}
\end{figure}

\begin{figure}[H]
\centering\includegraphics[width=0.75\textwidth]{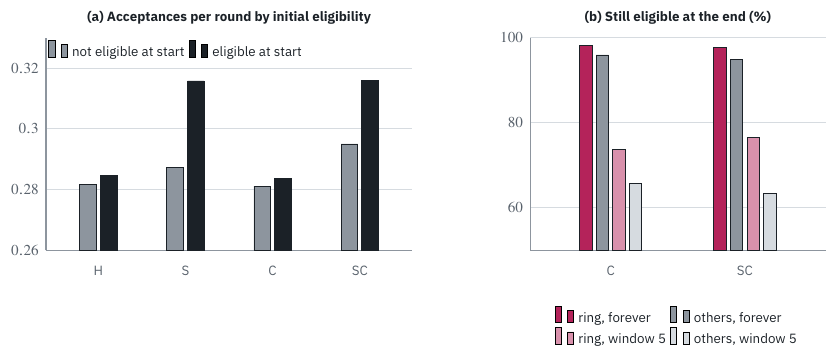}
\caption{\textbf{Eligibility loop.} ICLR-27-like rule, 500 agents, 20 rounds, 10 seeds. (a) Acceptances per round by eligibility at the start. (b) Share of agents still eligible at the end, ring members and others, with eligibility kept forever or for 5 rounds.}
\label{fig:eligibility}
\end{figure}

\section{Scale-Up Runs}\label{app:scaleup}

The scale-up ran \nscaleupWorlds{} conferences with seeds 1 to 5 on a second machine. Both machines reproduce each other's worlds bit for bit, so the scale-up and the original grids share seeds and are paired. Intervals are 95\% over seeds.

\begin{table}[H]
\centering\small
\caption{\textbf{Population size.} Accepted value relative to H with the same seed and population (\%), areas of 100, forward-looking agents, static budgets. The 500-agent column is the core grid with the same seeds.}
\label{tab:s1}
\begin{tabular}{lllrrr}
\toprule
Behaviour & Rule & Resources & 500 & 2{,}500 & 5{,}000\\
\midrule
S & Fixed & equal & $-7.0$ & $-6.4$ & $-6.4$ \\
S & Fixed & 80/20 & $-5.3$ & $-5.9$ & $-6.0$ \\
S & Mean & equal & $-13.4$ & $-13.8$ & $-13.1$ \\
S & Mean & 80/20 & $-5.9$ & $-7.0$ & $-6.7$ \\
C & Fixed & equal & $-0.2$ & $-0.3$ & $-0.2$ \\
C & Fixed & 80/20 & $-0.1$ & $-0.2$ & $-0.2$ \\
C & Mean & equal & $-0.2$ & $-0.1$ & $-0.1$ \\
C & Mean & 80/20 & $-0.3$ & $-0.2$ & $-0.3$ \\
SC & Fixed & equal & $-7.3$ & $-7.1$ & $-6.7$ \\
SC & Fixed & 80/20 & $-6.5$ & $-6.9$ & $-6.6$ \\
SC & Mean & equal & $-14.8$ & $-14.3$ & $-14.0$ \\
SC & Mean & 80/20 & $-7.3$ & $-7.8$ & $-7.3$ \\
 
\bottomrule
\end{tabular}
\end{table}

\begin{table}[H]
\centering\small
\caption{\textbf{Chosen alliances in pools of 500 and 1{,}000.} Fixed rule, equal budgets, detection on. ``Value'' is accepted value relative to H (\%). The prediction uses the saturating curve fitted on the \ncolNC{} cells of pools 50 to 250 (Figure~\ref{fig:dunbar}c). Out of sample, $R^2=\noosRsqC$ and RMSE \noosRMSEC{} for C, and RMSE \noosRMSESC{} with a mean error of $\noosBiasSC$ for SC.}
\label{tab:s2}
\begin{tabular}{lrrrrrrr}
\toprule
Behaviour & Pool $A$ & Cap & Allies & Value & $E$ & 1 $-$ overlap & Predicted\\
\midrule
C & 500 & 5 & 2.1 & $-0.1$ & 0.017 & 0.023 & 0.046 \\
C & 1,000 & 5 & 2.3 & $0.2$ & 0.009 & 0.013 & 0.040 \\
C & 500 & 15 & 7.7 & $-0.1$ & 0.062 & 0.070 & 0.079 \\
C & 1,000 & 15 & 7.3 & $-0.2$ & 0.029 & 0.038 & 0.055 \\
C & 500 & 50 & 31.0 & $-1.7$ & 0.249 & 0.201 & 0.194 \\
C & 1,000 & 50 & 30.8 & $-0.3$ & 0.123 & 0.120 & 0.120 \\
C & 500 & 150 & 105.3 & $-6.5$ & 0.844 & 0.360 & 0.408 \\
C & 1,000 & 150 & 102.7 & $-2.7$ & 0.411 & 0.269 & 0.271 \\
SC & 500 & 5 & 1.8 & $-6.4$ & 0.015 & 0.269 & 0.283 \\
SC & 1,000 & 5 & 1.9 & $-5.9$ & 0.008 & 0.262 & 0.280 \\
SC & 500 & 15 & 7.3 & $-6.0$ & 0.059 & 0.280 & 0.303 \\
SC & 1,000 & 15 & 6.6 & $-6.4$ & 0.026 & 0.267 & 0.289 \\
SC & 500 & 50 & 28.4 & $-7.1$ & 0.228 & 0.340 & 0.370 \\
SC & 1,000 & 50 & 29.4 & $-6.5$ & 0.118 & 0.296 & 0.329 \\
SC & 500 & 150 & 100.0 & $-12.9$ & 0.801 & 0.457 & 0.506 \\
SC & 1,000 & 150 & 94.8 & $-8.9$ & 0.379 & 0.380 & 0.418 \\
 
\bottomrule
\end{tabular}
\end{table}

\begin{table}[H]
\centering\small
\caption{\textbf{Larger surges.} ICLR-27-like rule with a 24-round eligibility window, equal budgets. Incumbent minus newcomer acceptance rate per paper (percentage points) in H and SC, and their paired difference. The 2{,}000 rows use seeds 1 to 5 of the original surge grid.}
\label{tab:s3}
\begin{tabular}{lrrrr}
\toprule
Growth & Final population & H & SC & SC $-$ H\\
\midrule
logistic & 2,000 & $0.0$ & $2.9$ & $2.9\pm0.4$ \\
logistic & 2,500 & $0.1$ & $2.4$ & $2.3\pm0.3$ \\
logistic & 5,000 & $0.4$ & $1.7$ & $1.4\pm0.5$ \\
exponential & 2,000 & $-0.1$ & $3.4$ & $3.5\pm0.7$ \\
exponential & 2,500 & $-0.1$ & $3.1$ & $3.2\pm0.6$ \\
exponential & 5,000 & $0.0$ & $2.6$ & $2.6\pm0.4$ \\
 
\bottomrule
\end{tabular}
\end{table}

\section{Language-Model Agents}\label{app:llm}

\paragraph{Protocol.} Each decision is one call with a system message that states the rules and a user message that states the agent's situation. The system message describes a repeated conference game with no fixed end (after every round it continues with probability 90\%), in which every researcher is both an author and a reviewer in the same area, so reviewed papers compete with the reviewer's own. It gives the score scale, the acceptance share and the submission rule. Authors are told the cost of each effort level and of fabrication, the detection probability per reviewer and the ban. Reviewers are told that they may report their honest score or, where allowed, a score two levels lower for a rival's paper or the top score for an ally's paper, and that any deviation is audited with probability 1\% and punished by a two-round suspension. Ring members are told that they belong to an informal alliance of five researchers. The stated goal is the agent's own acceptances, or those of the agent and its allies for ring reviewers in SC with the joint goal. The user message gives authors their budget, submission limit, last round's cutoff, record, and for each idea the cost, forecast mean score and acceptance chance of every affordable version. It gives reviewers the honest score, the allowed reports, last round's cutoff, their own submissions this round with forecast scores, and for ring members their allies' submissions. Replies are JSON objects. A reply that cannot be parsed is retried and then replaced by the numerical rule's choice. Worlds have 50 agents in one area, 10 rounds and two rings of five, with the calibrated parameters of the numerical model.

\paragraph{Models.} DeepSeek V4.1 Flash (\texttt{deepseek-flash}, low reasoning effort) through the official API, and Qwen3.7-Flash (\texttt{qwen3.7-flash}, thinking enabled with a budget of 1{,}024 tokens) through the international Model Studio endpoint. With Qwen, a thinking budget of 512 tokens gave fewer correct answers on a nine-question rule quiz (7.6 against 8.2). Replies that could not be parsed fell back to the numerical rule in \nllmDSFallback\% of DeepSeek decisions and in none of Qwen's. Qwen authors asked for more than their budget or submission limit allowed in \nllmQWTrimLo--\nllmQWTrimHi\% of decisions, and the simulator then dropped the last-listed papers until the choice fitted. Thinking budgets up to 4{,}096 tokens left this share above 40\%.

\begin{table}[H]
\centering\small
\caption{\textbf{Decisions by family and world}, seeds 1 to 5, mean $\pm$ 95\% interval over seeds (\%). Sabotage columns give the share of sabotage options taken. ``Ally boost'' is the share of ally reviews that give the top score. ``Fabricated'' is the share of submitted papers that are fabricated, and ``Submit nothing'' the share of author decisions with no submission.}
\label{tab:llm}
\begin{tabular}{llcccccc}
\toprule
 & & \multicolumn{3}{c}{Sabotage} & & & \\
\cmidrule(lr){3-5}
Family & World & S world & SC, outside rings & SC, ring members & Ally boost & Fabricated & Submit nothing\\
\midrule
DeepSeek & H Fixed & -- & -- & -- & -- & $0.7\pm0.6$ & $13\pm5$ \\
DeepSeek & H Mean & -- & -- & -- & -- & $0.4\pm0.7$ & $28\pm6$ \\
DeepSeek & S Fixed & $12.2\pm1.0$ & -- & -- & -- & $0.4\pm0.5$ & $7\pm3$ \\
DeepSeek & S Mean & $19.9\pm2.3$ & -- & -- & -- & $0.4\pm0.4$ & $16\pm3$ \\
DeepSeek & C Fixed & -- & -- & -- & $96.8\pm2.8$ & $0.7\pm0.7$ & $13\pm4$ \\
DeepSeek & C Mean & -- & -- & -- & $95.9\pm4.3$ & $0.2\pm0.3$ & $30\pm5$ \\
DeepSeek & SC Fixed & -- & $11.8\pm0.9$ & $75.2\pm2.1$ & $94.1\pm6.7$ & $0.2\pm0.4$ & $6\pm2$ \\
DeepSeek & SC Mean & -- & $19.7\pm1.5$ & $79.5\pm3.1$ & $95.4\pm5.4$ & $0.2\pm0.3$ & $13\pm3$ \\
DeepSeek & SC Fixed, own goal & -- & $11.6\pm1.6$ & $39.4\pm7.0$ & $94.6\pm6.4$ & $0.5\pm0.5$ & $7\pm3$ \\
Qwen & S Fixed & $13.8\pm0.6$ & -- & -- & -- & $55.4\pm2.4$ & $2\pm2$ \\
Qwen & S Mean & $9.8\pm1.5$ & -- & -- & -- & $54.3\pm1.9$ & $1\pm1$ \\
Qwen & SC Fixed & -- & $13.2\pm1.0$ & $30.0\pm2.8$ & $96.4\pm5.1$ & $56.1\pm4.2$ & $2\pm2$ \\
Qwen & SC Mean & -- & $11.2\pm1.0$ & $32.4\pm2.0$ & $97.9\pm5.8$ & $51.3\pm1.4$ & $2\pm2$ \\
Qwen & SC Fixed, own goal & -- & $14.5\pm1.8$ & $26.3\pm2.4$ & $95.5\pm9.0$ & $55.2\pm1.0$ & $2\pm1$ \\
 
\bottomrule
\end{tabular}
\end{table}

\begin{table}[H]
\centering\small
\caption{\textbf{DeepSeek alliance worlds}, seeds 1 to 5, mean $\pm$ 95\% interval. Standing is the percentile of chosen partners in past acceptances (0.5 is random), for the language model and for the numerical rule shown the same choices. Accepted and renewed are shares of proposals accepted and of ties proposed again in the next round (\%). ``Pacts kept'' is the share of ally reviews that give the top score (\%). Allies are realised ties per agent in late rounds.}
\label{tab:ally}
\begin{tabular}{lcccccc}
\toprule
Rule, resources & Standing (LLM) & Standing (numerical) & Accepted & Renewed & Pacts kept & Allies\\
\midrule
Fixed, equal & $0.45\pm0.03$ & $0.34\pm0.04$ & $21\pm3$ & $57\pm4$ & $86\pm9$ & $2.2\pm0.1$ \\
Fixed, 80/20 & $0.46\pm0.03$ & $0.32\pm0.03$ & $21\pm7$ & $58\pm3$ & $85\pm6$ & $2.1\pm0.2$ \\
Mean, equal & $0.55\pm0.04$ & $0.34\pm0.02$ & $18\pm3$ & $58\pm2$ & $77\pm5$ & $2.1\pm0.1$ \\
Mean, 80/20 & $0.57\pm0.05$ & $0.33\pm0.02$ & $19\pm3$ & $58\pm2$ & $80\pm8$ & $2.2\pm0.2$ \\
 
\bottomrule
\end{tabular}
\end{table}

\begin{table}[H]
\centering\small
\caption{\textbf{DeepSeek world outcomes.} Accepted value relative to the language-model H world with the same seed and rule (\%), seeds 1 to 5, mean $\pm$ 95\% interval.}
\label{tab:llmvalue}
\begin{tabular}{llr}
\toprule
Behaviour & Rule & Value vs H\\
\midrule
S & Fixed & $+8.4\pm3.9$ \\
C & Fixed & $0.0\pm1.7$ \\
SC & Fixed & $+4.2\pm7.5$ \\
S & Mean & $+15.3\pm7.8$ \\
C & Mean & $-3.2\pm4.3$ \\
SC & Mean & $+17.5\pm4.4$ \\
 
\bottomrule
\end{tabular}
\end{table}

\fi

\end{document}